\documentclass[fleqn,10pt]{wlscirep}
\usepackage[utf8]{inputenc}
\usepackage[T1]{fontenc}
\usepackage{adjustbox}
\usepackage{graphicx,subcaption,booktabs,siunitx}
\usepackage{multirow}
\usepackage{soul}

\title{From Toxicity to Conformity: Adaptive user behavior to social norms in Telegram communities}

\author[1,2,+]{Lorenzo Alvisi}
\author[1,3,+,*]{Victoria Popa}
\author[1]{Guglielmo Cola}
\author[1]{Serena Tardelli}
\author[1]{Maurizio Tesconi}
\affil[1]{Institute of Informatics and Telematics, National Research Council, Pisa, Italy}
\affil[2]{IMT School for Advanced Studies Lucca, Lucca, Italy}
\affil[3]{University of Pisa, Computer Science Department, Pisa, Italy}

\affil[*]{email: victoria.popa@phd.unipi.it}

\affil[+]{these authors contributed equally to this work.}

\begin{abstract}
 
Toxic and antisocial user behavior on social media platforms has received considerable scholarly attention due to its detrimental effects on society.
This study takes a holistic perspective on the phenomenon of online toxicity by investigating the impact of local community norms on toxic expression. By using six large-scale datasets, comprising over 500 million Telegram messages collected between 2015 and 2024, we analyze toxic user behavior across multiple chats and languages. We introduce a methodological framework that models user adaptation through a conformity index, capturing conformist, anti-conformist, and independent behavioral tendencies. Our findings show that most users tend to conform to local normative environments, adjusting their toxicity to match the toxicity levels of the chats in which they participate. These patterns are consistent across datasets and languages, suggesting that community norms and social influence play a decisive role in shaping user behavior online. Furthermore, we demonstrate that exposure to these norms, in terms of increased user participation in chats, is associated with a stronger tendency toward conformity with the surrounding social contexts. Collectively, these findings contribute to a deeper understanding of toxic online behavior and highlight the importance of contextualized approaches to content moderation.

\end{abstract}
\begin{document}

\flushbottom
\maketitle

\thispagestyle{empty}

\section*{Introduction}
Human behavior, to be comprehensively understood, must be examined in relation to the social contexts in which it occurs. Individuals do not act in isolation; instead, their opinions and actions are shaped by the surrounding environment, interpersonal relationships, and social norms~\cite{cialdini2004social}. Consequently, the context in which people interact exerts a substantial influence on their behavior. This principle, referred to as “situational power”, emphasizes the critical role of external factors and the surrounding social environment in shaping individuals' thoughts, emotions, and actions~\cite{ross2011person, pryor2019even}. Since social interaction is mirrored in digital spaces, research has focused on how offline behavioral dynamics translate online, where social media platforms exert a strong influence on the shaping of perceived norms~\cite{centola2015spontaneous, masur2021behavioral}. As a result, studies show that social norms play a powerful yet double-edged role in shaping behavior and interaction within online communities~\cite{matias2019preventing, chandrasekharan2017you}. While they can promote positive outcomes such as social cohesion and collective activism~\cite{greijdanus2020psychology}, they can also contribute to polarization, exclusion, and the spread of toxic~\cite{shen2020viral, masur2021behavioral, garimella2018political, cinelli2021echo} and conspiratorial behaviors~\cite{senette2025unpacking,gambini2024anatomy, alvisi2024unraveling}.
Therefore, even online toxicity, as a human product, cannot be fully comprehended or examined in isolation from its contextual setting. Instead, it can provide a valuable lens through which to gain insight into the underlying social norms and identity dynamics that characterize online communities.

Toxicity in digital spaces is a prominent research topic, as detrimental behaviors such as harassment, hate speech, and incivility are pervasive across social media platforms, and increasingly affect the tone and quality of public conversations, degrading the health of online communities and the overall user experience~\cite{castano2021internet}. Despite substantial academic and applied interest in the topic, the state of the art still faces significant challenges in defining online toxicity with precision, which in turn hinders its effective detection and mitigation through robust moderation strategies~\cite{sheth2022defining, fortuna2020toxic}. These issues stem from the absence of a universally agreed conceptual framework, leading to the broad categorization of various antisocial behaviors, such as incivility~\cite{borah2014does}, hate speech~\cite{castano2021internet}, cyberbullying~\cite{hosseinmardi2015analyzing, giumetti2022cyberbullying}, and harassment~\cite{gagliardone2015countering, castano2021internet} under the general umbrella label of ``toxicity.'' 
In light of these challenges, online toxicity is generally measured using the Perspective API, a validated classifier developed by Google's Jigsaw~\cite{lees2022new} and widely used in the literature for the automated detection of toxic speech.
Numerous studies use the Perspective API toxicity score to examine online toxicity and antisocial behavior on social media platforms~\cite{xia2020exploring, mamakos2023social, avalle2024persistent, blumer2025tracking}.

Classical approaches in this area have focused mainly on the detection and mitigation of toxic behavior in online environments, with a strong emphasis on the analysis of text-level content~\cite{xia2020exploring, avalle2024persistent, cinelli2021dynamics, saveski2021structure}. This method, while valid, can often overlook the emerging social dynamics within online communities that contribute to the persistence of toxicity on a broader scale. Sheth et al. (2022)~\cite{sheth2022defining} highlight the limits of current approaches to identifying and addressing online toxicity, noting that these methods often focus narrowly on isolated posts or content, disregarding the broader systemic and contextual factors in which such toxicity occurs.
In fact, further research has extended the focus to community-level dynamics~\cite{almerekhi2022investigating, cheng2015antisocial}, shedding light on how community norms~\cite{saveski2021structure, cheng2017anyone}, platforms' architecture~\cite{munn2020angry}, and content moderation policies~\cite{blackwell2017classification, cima2025investigating} can foster toxicity diffusion.
Other studies have explored toxicity from a longitudinal perspective, examining how individual user behavior evolves over time and across communities, with the aim of characterizing the behavioral patterns associated with toxic engagement~\cite{blumer2025tracking, mall2020four}. 
In general, the role of toxicity as a social norm and the normalization process of online toxic behavior have been explored most extensively within specific contexts such as online gaming communities, where toxic behaviors are more easily accepted and often embedded in group culture. Studies show that players may view toxic behavior as normalized~\cite{beres2021don} and that exposure to it increases the likelihood of imitation and contagion~\cite{shen2020viral}.
Among the studies that explored user behavior and the influence of context, Rajadesingan et. al (2020)~\cite{rajadesingan2020quick} analyzed users on politically oriented subreddits on the Reddit platform and demonstrated that they quickly adapt to the specific community toxicity norms and do not carry over their behavior from one community to another.  
This finding is further supported by a previous study by Chandrasekharan et al. (2017)~\cite{chandrasekharan2017you} which showed that after Reddit banned hate subreddits, members of those places did not engage in similar hate speech in other subreddits they joined later.
This suggests that user behavior is context-sensitive and shaped by local community norms and that behavioral contagion may be localized in specific context but not exported outside.

Existing research on online toxicity has predominantly focused on platforms such as Reddit and X (formerly Twitter), while the Telegram environment remains comparatively underexplored, despite its considerable potential for advancing understanding of these dynamics. In fact, where Telegram has been studied, toxicity has often appeared as a marginal concern within broader investigations of conspiracy communities and extremist discourse~\cite{hoseini2023globalization}. Telegram's platform structure, organized around communities in the form of chats, enables the observation of individual user behavior in different communities without the confounding influence of algorithmic curation, which can otherwise amplify or promote toxic dynamics. Within this framework, this social messaging platform represents a particularly relevant case for studying online social dynamics. Unlike other social platforms, it is characterized by the absence of recommendation algorithms and by a limited search function, restricting user exposure to external norms and influences while encouraging the emergence of relatively closed and autonomous communities. These features make Telegram a valuable setting for observing how social and identity norms emerge and naturally consolidate in digital environments. This, in turn, provides a useful vantage point for investigating the development and persistence of toxic behaviors online.

In summary, although online toxicity has been extensively studied, several gaps are present in the literature. In particular, much of the research focuses on isolated cases of highly toxic user behavior and on the importance of content moderation, with limited attention to how toxicity is influenced by context and local community norms. Studies addressing the social dimension of toxicity often examine specific contexts, such as gaming, without considering how individuals adapt their behavior across different communities and settings. In addition, longitudinal and cross-community perspectives remain comparatively underexplored. To contribute to addressing these gaps, we adopt a multidisciplinary and context-sensitive approach using Telegram data to examine how users' toxic behavior aligns with or deviates from the normative pressure of different communities. 
Specifically, we explore and quantify users' toxic behavior across multiple communities in which they actively participate, accounting for both individual variations in toxicity and the overall toxicity levels of each community, leveraging six different datasets in multiple languages, for a total of around 500 million messages. In addition, this work contributes a large-scale dataset containing over 200 million messages collected from Telegram throughout 2024.

We show that, at a global level, users tend to adjust their behavior in line with the context in which they participate, aligning with the prevailing behavioral norms in the chat. These dynamics are consistently observed across all the analyzed datasets, highlighting the strong influence of local community norms on user behavior in multilingual settings.
In addition, we analyze individual user behavior in relation to their surrounding context and identify three types of behavioral patterns: conformist, anti-conformist, and independent of local community norms. Among the three behavioral types, we find that conformist users represent the majority compared to anti-conformists and independents, indicating that adaptation to local norms is the dominant tendency in the studied communities.

\section*{Results}

\subsection*{Preliminaries}

\subsubsection*{Datasets}
Our analysis draws on six Telegram datasets that cover various contexts and languages. This diversity enables the examination of whether user adaptation and toxic behavior are consistent across different settings. Specifically, we use one dataset previously published in the literature~\cite{alvisi2025mapping}, one large multilingual dataset already published~\cite{baumgartner2020pushshift} that we subdivide into four language-specific subsets (English, Italian, Russian, and Portuguese), and one multilingual dataset we created for this study. These datasets show minor differences in data collection methodologies, as described in Section Methods. A detailed summary of dataset sizes is provided in Table~\ref{tab:datasets_ccdf}.

\paragraph{User selection}
We consider user-chat pairs in which the user has contributed at least one hundred messages, and we further restrict the dataset to users who satisfy this condition in at least two distinct chats. This criterion ensures that we focus on active participants and allows us to capture changes in their behavior across different interaction contexts. Further details on the processed datasets are provided in Section Methods.

\subsubsection*{Toxicity metrics}
To analyze expressions of toxicity and users' behavioral shifts, we rely on a set of metrics. We employ the Perspective API~\cite{lees2022new}, a state-of-the-art validated tool for detecting toxic language. The API assigns each Telegram message a toxicity score between 0 and 1, with messages classified as toxic if the score exceeds a predefined threshold, as suggested by Perspective API documentation.

\paragraph{Chat toxicity}
The toxicity of a chat is defined as the percentage of toxic messages within a given chat.

\paragraph{User toxicity}
The toxicity of a user in a chat is defined as the percentage of toxic messages posted by a user within a specific chat.

\begin{figure}[t!]
\centering
\includegraphics[width=0.4\linewidth]{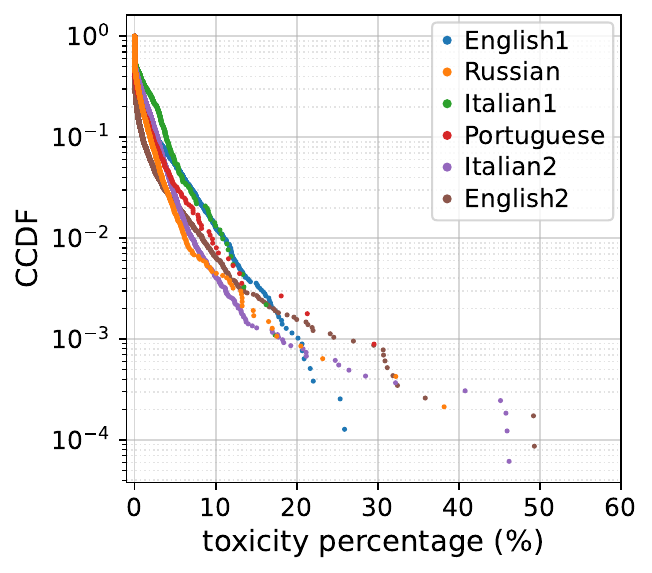}
\caption{Complementary cumulative distribution function (CCDF) of chat toxicity across datasets.
All datasets, despite linguistic and contextual differences, display similar tail behavior indicating the presence of both low-toxicity and highly toxic communities, providing variability for cross-environment behavioral analysis.} 
\label{fig:ccdf}
\end{figure}

\paragraph{Chat toxicity distribution} 
In Figure~\ref{fig:ccdf} we report the complementary cumulative distribution function (CCDF) of chat toxicity. All datasets display a clear tail behavior, either approximately exponential or heavier-than-exponential, consistent with the presence of long- or short-tailed regimes. This observation confirms that each dataset includes different levels of chat toxicity, an essential condition for examining changes in user behavior across different environments.

\subsection*{Global patterns of adaptive user behavior}
\begin{figure}[t]
  \centering
s  \begin{subfigure}[t]{0.49\textwidth}
    \includegraphics[width=\linewidth]{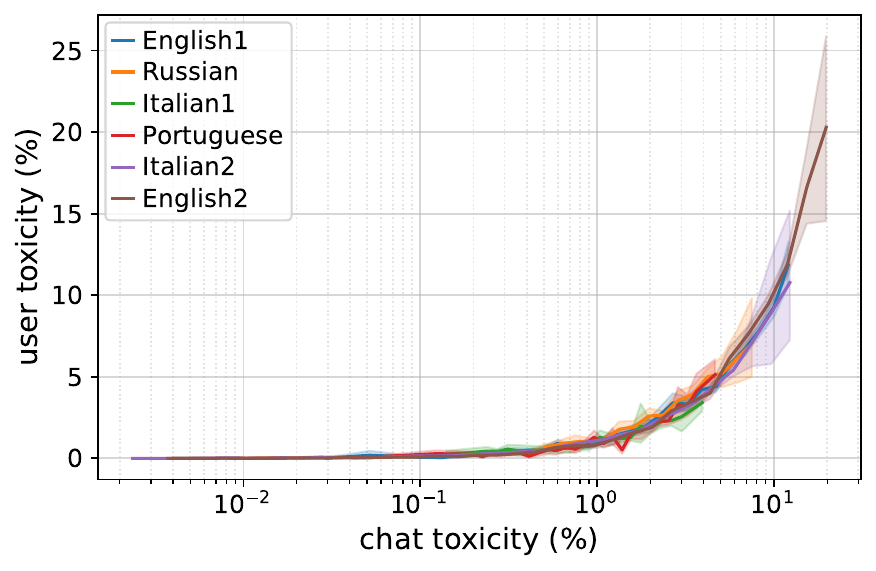}
    \subcaption{} 
  \end{subfigure}\hfill
  \begin{subfigure}[t]{0.49\textwidth}
    \includegraphics[width=\linewidth]{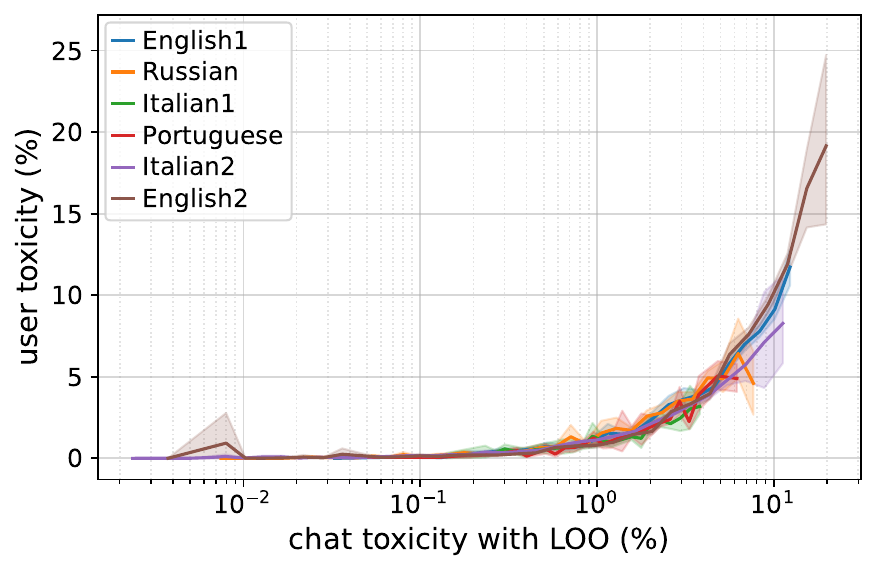}
    \subcaption{}
  \end{subfigure}

  \caption{Correlation between user toxicity and chat toxicity. Panel (a) shows that, across all datasets, user toxicity increases with the toxicity level of the chat they participate in. The same pattern holds in Panel (b) when removing the user's own contributions (Leave-One-Out), confirming that the effect reflects genuine environmental influence. Shaded areas represent 95\% confidence intervals. Correlation values are reported in Table~\ref{tab:combined_corr}.}
  
  \label{fig:q1}
\end{figure}

Our first research question investigates whether Telegram users collectively adapt their language and tone to the environments in which they interact, aligning with the prevailing toxicity levels of the chats in which they participate.
In fact, we assume that context shapes user behavior and hypothesize that users adapt their behavior according to the chats they engage in.
To test this hypothesis, we assess the extent of such adaptation by computing the correlation between chat toxicity and user toxicity (Figure~\ref{fig:q1}a).
Since chat toxicity distributions display distinct tail behavior, we apply logarithmic binning to the data. This approach reduces noise and local fluctuations, enabling clearer visualization of global trends that are otherwise difficult to observe in the raw distribution. The effectiveness of this methodology is well supported and validated in the literature~\cite{milojevic2010power,labo2024using}. As shown in Figure~\ref{fig:q1}a, all datasets display a common increasing trend. We observe strong correlations between user toxicity and chat toxicity across all datasets and languages considered, indicating that users tend to be more toxic in chats characterized by higher overall toxicity levels. 
To further validate and strengthen this result, we implement a robustness check based on a leave-one-out framework. For this analysis, we consider the user's toxicity in a chat as the percentage of toxic messages within that chat, excluding the user's own contributions. As such, the estimated chat toxicity captures only the surrounding environment and avoids user bias. As shown in Figure~\ref{fig:q1}b, the consistency of the results across definitions further supports the conclusion that the observed association is not an artifact of the chosen metric. 

\begin{table}[t]
\centering
\begin{tabular}{lcccrr}
\toprule
 & \multicolumn{3}{c}{\textbf{Correlations}} & \multicolumn{2}{c}{\textbf{Baselines}} \\
\cmidrule(lr){2-4} \cmidrule(lr){5-6} 
\textbf{Dataset} 
& \begin{tabular}[c]{@{}c@{}}Chat--User\\ (Figure~\ref{fig:q1}a)\end{tabular} 
& \begin{tabular}[c]{@{}c@{}}LOO\\ (Figure~\ref{fig:q1}b)\end{tabular}  
& \begin{tabular}[c]{@{}c@{}}Dynamic\\ (Figure~\ref{fig:delta})\end{tabular}  
& \multicolumn{1}{c}{\begin{tabular}[c]{@{}c@{}}Random\\ (Figure~\ref{fig:q1}a)\end{tabular}}
& \multicolumn{1}{c}{\begin{tabular}[c]{@{}c@{}}Random\\ (Figure~\ref{fig:delta})\end{tabular}} \\

\midrule
English1   & $0.998$** & $0.996$** & $0.866$** & $ 0.014 \pm 0.109$ & $-0.053 \pm 0.373$ \\
Russian    & $0.999$** & $0.952$** & $0.831$** & $-0.057 \pm 0.284$ & $-0.015 \pm 0.290$ \\
Italian1   & $0.988$** & $0.980$** & $0.820$** & $-0.019 \pm 0.218$ & $ 0.001 \pm 0.306$ \\
Portuguese & $0.983$** & $0.974$** & $0.950$** & $-0.008 \pm 0.173$ & $ 0.006 \pm 0.277$ \\
Italian2   & $0.999$** & $0.994$** & $0.928$** & $-0.136 \pm 0.386$ & $-0.040 \pm 0.409$ \\
English2   & $0.999$** & $0.998$** & $0.993$** & $ 0.000 \pm 0.230$ & $ 0.001 \pm 0.280$ \\
\bottomrule
\end{tabular}
\caption{Correlations and baselines across datasets.
Correlations measure the relationship between user and chat toxicity: 
(a) standard chat--user correlation (Figure~\ref{fig:q1}a), 
(b) leave-one-out correlation excluding the user's own messages (Figure~\ref{fig:q1}b), 
and ($\Delta$) dynamic correlation between changes in user and chat toxicity (Figure~\ref{fig:delta}). 
Baselines are computed by randomizing user--chat associations. 
Double asterisks (**) denote statistical significance ($p < 0.01$).}
\label{tab:combined_corr}
\end{table}

\begin{figure}[t]
  \centering
  \includegraphics[width=0.50\linewidth]{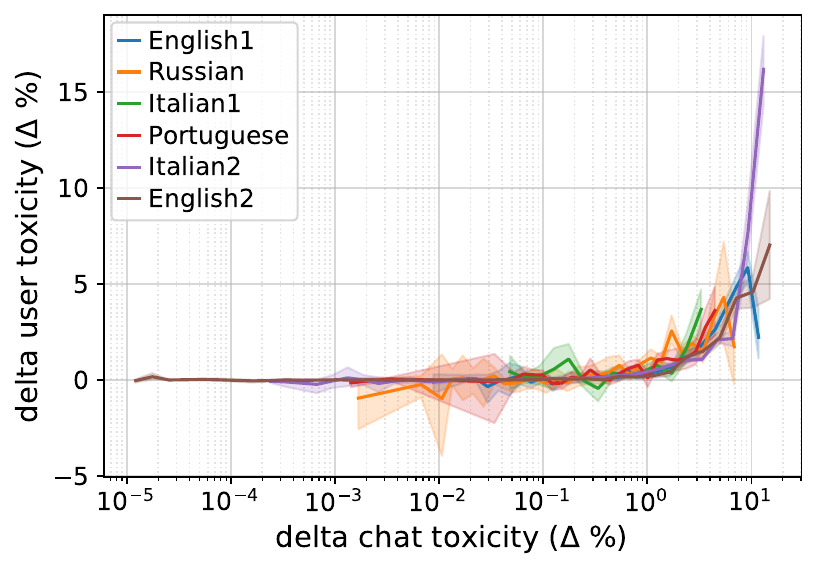}
  \caption{Change in user toxicity as a function of environmental shifts. For every user active in at least two chats, the plot shows how changes in community toxicity correspond to changes in the user's own toxicity. All datasets display a clear positive trend: users become increasingly toxic when moving into more toxic environments. This dynamic evidence shows that user behavior is shaped by contextual norms rather than static personal tendencies.}
  \label{fig:delta}
\end{figure}

The previous analyses do not take into account whether the same user engages in environments with different levels of toxicity. 
To account for this, we test whether users adapt their behavior to the surrounding environment, rather than simply self-selecting into chats with toxicity levels they find acceptable. Therefore, for each pair of chats in which a user participates, we compute two quantities: the variation in environmental toxicity and the corresponding change in the user's behavior. Specifically, we define the variation in environmental toxicity as the difference between the toxicity of the more toxic chat and that of the less toxic one, which is therefore always non-negative. In contrast, the change in user behavior is defined as the difference in the user's own toxicity between the more toxic and less toxic chat, and can take values in the range $[-1,1]$. To reduce the noise, we applied the same binning methodology as in the previous analysis. The results, presented in Figure~\ref{fig:delta}, show a consistent trend  and strong correlations across all datasets, supported by statistically significant p-values. This analysis further implies that users' use of toxic language is strongly correlated with environmental conditions, showing that users not only behave in line with those conditions but also change their behavior in response to them.

Finally, we test the robustness of these findings by randomizing the chat toxicity values and repeating the entire analysis 1,000 times. Randomized analyses yield non-significant correlations, and the resulting toxicity distributions appear to be uniform, producing a flat trend. We further test the reliability of the results of both Figure~\ref{fig:q1}a and Figure~\ref{fig:delta} with respect to variations in the number of bins and threshold values, demonstrating that the observed patterns remain stable even across different configurations.

\subsection*{Individual patterns of user norm conformity}

We then aim to investigate whether, and to what extent, individual users consistently adapt their behavior and conform to environmental norms.
We acknowledge that, due to the ecological fallacy~\cite{piantadosi1988ecological}, aggregate patterns cannot be used to infer individual behavioral tendencies. Therefore, even though we observed consistent global patterns and user adjustments to local normative contexts, we cannot assume that individual users necessarily exhibit adaptive behavior. 
Consequently, to analyze individual behavior, we introduce an analytical framework that captures how users adjust their toxicity in response to the environmental level of toxicity within chats. 

\begin{figure}[t]
    \centering
  \begin{subfigure}[t]{0.38\textwidth}
    \includegraphics[width=\linewidth]{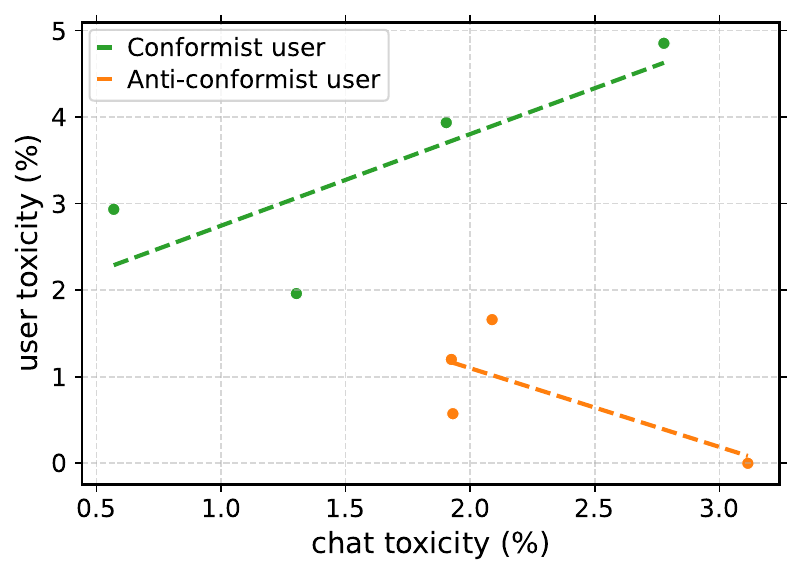}
    \subcaption{} 
  \end{subfigure}
  \begin{subfigure}[t]{0.48\textwidth}
    \includegraphics[width=\linewidth]{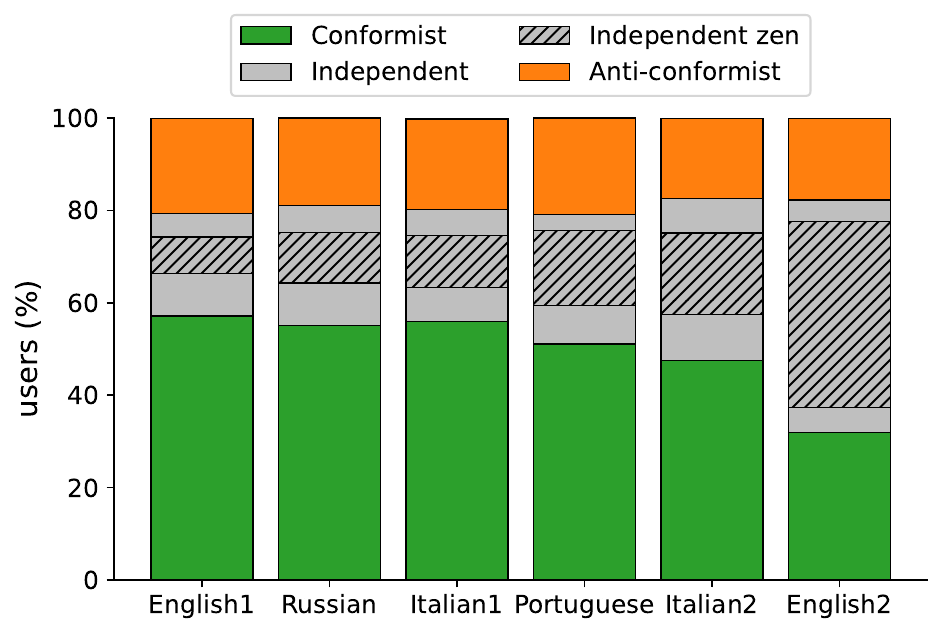}
    \subcaption{}
  \end{subfigure}

  \vspace{0.8em}

  \begin{subfigure}[c]{\textwidth}
    \includegraphics[width=\linewidth]{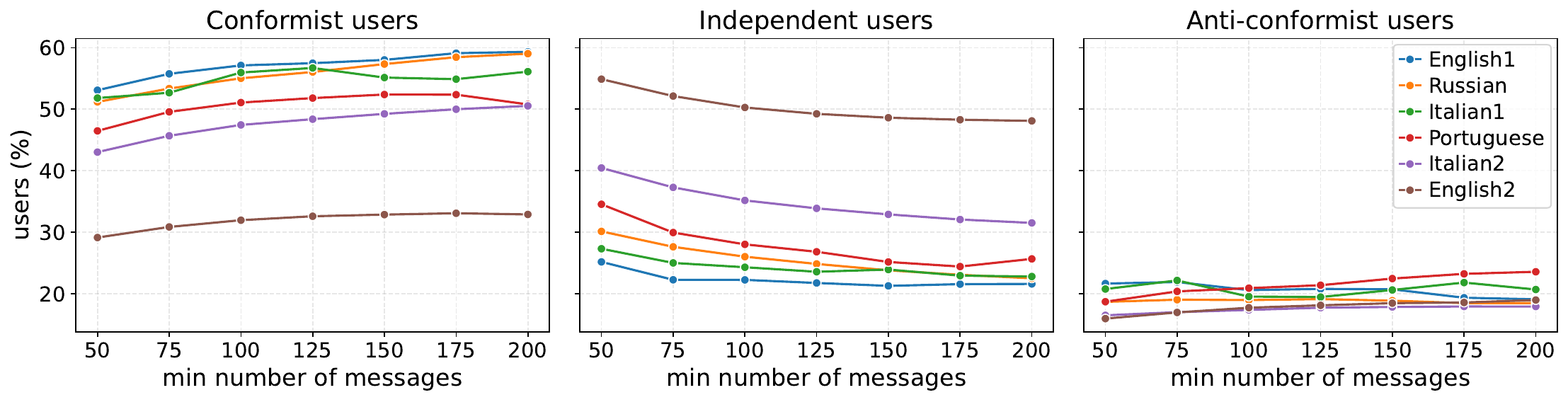}
    \subcaption{}
  \end{subfigure}\hfill

  \caption{Individual patterns of conformity to community norms: 
  Panel (a) shows two examples of fitted user trends, showing how individual user-specific slopes capture different adaptive behavior to chat environments.
  Panel (b) shows the distribution of these behavioral types across datasets, where the majority of users appear to be conformists, with independents and anti-conformists representing smaller shares. This pattern remains consistent when tested across different slope thresholds (Figure~\ref{fig:thresh-origin-label}).
  Panel (c) shows how these proportions change when increasing the minimum number of messages required for inclusion, revealing that higher user activity and engagement reduces independence and reinforces conformity, as increased exposure to community norms reinforces norm alignment.}
  \label{fig:q2}
\end{figure}

For each user, we consider all chats in which they have posted, computing both the local environmental toxicity and the degree of toxic language they exhibit within that context. We fit a regression line through these points, extracting a slope and an intercept for each user. 
We conceptualize these slopes as a \textit{conformity index}, capturing three types of user responses to contextual toxicity: \textit{conformist} (positive slopes, indicating user adaptation to the environment), \textit{anti-conformist} (negative slopes, indicating an opposite adjustment), and \textit{independent} (slopes near zero, indicating behavioral stability across different contexts, and adherence to a consistent behavior). Another interesting finding emerges within the independent category. Notably, we identify a group of users who exhibit no toxicity in any context and consistently refrain from using toxic language. We refer to them as \textit{zen users} to emphasize their consistent self-control. 
In summary, this methodology allows for the identification and classification of different behaviors emerging from observable data. The example in Figure~\ref{fig:q2}a shows how the resulting slopes may vary between users, revealing heterogeneous and context-dependent online behavioral patterns.

In formalizing this methodological framework, we anchor it to existing theoretical work. In particular, our approach resonates with classical models of social psychology, and specifically with Willis's tripartite model of social response~\cite{willis1963two}. This model considers conformity, independence, and anti-conformity as distinct dimensions of individual reactions to social influence. Although more recent and comprehensive psychological models also account for the interplay between public and private spheres in explaining individual conformity pressure~\cite{nail2013proposal}, the limitations of our dataset constrain our analysis to observable public behaviors within Telegram chats. The alignment between our empirical framework and Willis's model provides a strong theoretical foundation to understand the behavioral patterns observed in our data.
This correspondence supports our interpretation that user behavior is molded by contextual factors, with local community normative rules exerting a significant influence in shaping user conduct.

Given the heterogeneous behaviors observed among users, we further examine their predominant tendencies and distribution. Figure~\ref{fig:q2}b shows the distribution of behavioral categories across datasets: in all cases, most users exhibit conformist behavior, adapting their toxicity to match the surrounding environment. A smaller proportion of users can be classified as anti-conformist, whereas the share of independent users varies more substantially across datasets, reaching particularly high values in \texttt{English2}. We attribute this pattern to the dataset's collection methodology and its thematic focus on cryptocurrencies, as opposed to the broader topical coverage characterizing the other datasets.

Another important aspect we aim to investigate concerns how users' behavior varies with respect to the number of messages they sent in chats. As outlined in \textit{Preliminaries}, our analysis considers users who posted at least one hundred messages across a minimum of two different chats each. Therefore, we study how the results in Figure~\ref{fig:q2}b change as the minimum threshold of messages sent by users in chats varies. 
Figure~\ref{fig:q2}c shows how the distribution of users across the three categories: conformists, anti-conformists, and independents changes as the minimum chat activity threshold increases from 50 to 200 messages. As thresholds rise, the share of independent users decreases, anti-conformists remain stable, and conformists become more prevalent. This indicates that greater platform usage, together with an increased exposure to community norms, reduces independence while reinforcing conformity and contextual adaptation.

\section*{Discussion} 

In this study, we show that users generally tend to adjust their behavior to the specific contexts in which they interact, thus conforming to the local norms of the online community with regard to toxic behavior. Therefore, we can assume that context plays a decisive role in shaping individuals' toxic behavior.
This pattern emerges consistently from both the global and individual-level analyses of users within the Telegram communities investigated, spanning all datasets and languages considered in this work.
In fact, the global analysis shows a strong relationship between user toxicity and chat toxicity, as shown in Figure~\ref{fig:q1}a. This finding is further supported by the analysis in Figure~\ref{fig:q1}b, which demonstrates that the relationship persists even after excluding biases arising from the user's own contributions.
Even when focusing on individual-level behavior, the analysis reveals interesting patterns. As shown in Figure~\ref{fig:q2}b, individual users tend to adjust their conduct in response to varying contextual conditions. What emerges is that across all datasets most online users tend to adopt conformist behaviors relative to the context, while anti-conformist and independent tendencies are present but less prevalent, with the exception of the \texttt{English2} dataset.
Furthermore, Figure~\ref{fig:q2}c highlights another important result: the more users interact online, the more likely they are to conform to their reference environment and adapt to the prevailing social norms of the community. Notably, while the proportion of anti-conformist users remains relatively stable even under increased exposure to community norms, the share of independent users decreases and appears to merge numerically into the conformist category.
These findings suggest that online toxicity may not necessarily be the outcome of antisocial personality traits, although they can contribute to amplifying it; rather it appears as a large-scale, context-dependent social phenomenon shaped by community norms, conformity pressures, and group dynamics.
Therefore, toxicity can serve as a valuable proxy for identifying social norms within communities, helping to delineate what is considered acceptable or unacceptable in specific groups. Moreover, in some contexts where toxicity becomes normalized, such as certain gaming communities, and is internalized by users, it may evolve into a social norm in its own right.

Our findings contribute to a growing understanding of online user toxic behavior as a complex social phenomenon shaped not only by individual dispositions but also by the interplay of contextual, relational and structural factors. 
This perspective aligns closely with insights from social psychology, particularly theories of normative influence and individual behavioral adaptation to conformity mechanisms.
In fact, a substantial body of research has examined the dynamics of individual conformity to group norms and expectations~\cite{sherif1936psychology, asch1956studies, deutsch1955study, cialdini2004social, reno1993transsituational}. As demonstrated in Asch (1956)\cite{asch1956studies} conformity experiments, collective pressure has the power to lead individuals to align themselves with the majority opinion, even though there is no real internal adherence. In fact, the desire to avoid disapproval or social exclusion appears to prevail more over personal judgment, highlighting the powerful role of normative influence~\cite{deutsch1955study} in shaping individual behavior even in situations where it is recognized that these norms are arbitrary~\cite{pryor2019even} or do not align with the actual beliefs of individuals~\cite{asch1956studies, kelman1958compliance}. 
This theory may provide a plausible explanation for why users tend to adapt their behavior to the specific communities they join.

In addition, the structural features of online platforms appear to play a crucial role in shaping these dynamics and spreading individual toxicity.
The General Aggression Model (GAM)~\cite{allen2018general} highlights how individual traits and situational factors are crucial in influencing aggressive behavior.
Features such as anonymity, disembodiment, and reduced accountability amplify the spread of toxicity by fostering  disinhibition in user behavior~\cite{pandita2024three}. 
Moreover, according to the Identity Model of Deindividuation Effects (SIDE)\cite{reicher1995social}, visual anonymity enhances the salience of group social identities and individual conformity to group normative behavior. Consequently, individuals are more susceptible to group influence when they are anonymous than when they are personally identifiable~\cite{postmes2000formation}. In this regard, Telegram provides a high degree of anonymity, reinforcing the conditions under which such dynamics can emerge.
In this sense, online platforms do not merely host toxic interactions, but may actively contribute to creating conditions conducive to the spread of toxicity~\cite{cinelli2021echo}. Most studies on this research area have mainly explored platforms such as X (formerly Twitter) and Reddit, which have structural and algorithmic constraints that tend to obscure community-level interactions and make it difficult to observe behavioral patterns independent of algorithmic intervention~\cite{gillespie2018custodians, devito2017algorithms}.
By conducting these analyses on Telegram, we were able to observe behavioral dynamics and shifts in user toxicity within communities with greater transparency.

Our findings highlight that toxicity in online spaces is not driven solely by a handful of severely toxic users who distort community discourse~\cite{cheng2015antisocial, jhaver2019does}, but is embedded in collective and contextual dynamics that reinforce conformity to local norms. This perspective challenges the dominant intervention strategies, which have often prioritized the detection and removal of toxic content or individuals, and instead emphasizes the importance of addressing the systemic and community-level conditions that allow toxicity to persist~\cite{gorwa2020algorithmic}. Consequently, moderation processes should also account for the toxicity of the context and communities in which users interact, as they represent an influential factor in shaping behavior.

We acknowledge that our study presents some limitations. First, although our analyses are based on a large volume of user data, they are restricted to a single platform, namely Telegram. In particular, Telegram's structural features, such as large group sizes, weak moderation, and limited accountability, may facilitate the persistence of toxic behavior by enabling harmful norms to become embedded within communities. As such, we need to be cautious in generalizing our findings to other online environments, where platform design and moderation policies may shape user behavior differently. Future work should extend this analysis to other platforms to assess the extent to which these findings hold across different platforms.
Furthermore, there are some limitations related to the datasets that should be highlighted. First, the datasets are multilingual, and Perspective API, despite its level of accuracy, may still not fully capture cultural and linguistic nuances, causing potential errors in toxicity classification. Second, the selected datasets may not be representative of Telegram communities in general.
Moreover, since the study is observational and correlational, it is not possible to determine to what extent contextual factors, such as chat topics, directly influence users' behavior. Future research should explore how specific community dynamics influence toxic norms.

\section*{Methods}

\subsection*{Toxicity detection}
The definition of online toxicity is a well-recognized challenge, as it is often used as an umbrella term that encompasses various forms of antisocial communication~\cite{avalle2024persistent,sap2021annotators,pavlopoulos2020toxicity}. For this reason, researchers often rely on an operational definition, such as the one adopted by Perspective API\cite{lees2022new}, which defines toxicity as ``a rude, disrespectful, or unreasonable comment that is likely to make you leave a discussion''~\cite{lees2022new,salminen2020topic,avalle2024persistent,blumer2025tracking}. Perspective API, developed by the Google Jigsaw team, assigns to each textual element a continuous score between 0 and 1. This score represents the estimated probability that a typical reader would perceive the text as toxic. Following previous works~\cite{alvisi2025mapping,avalle2024persistent,bruni2025amaqa}, we consider a comment toxic when it has a toxicity score greater than or equal to 0.7, the threshold recommended by the official documentation. This implies that approximately 70\% of the readers would likely perceive the comment as toxic.

\subsection*{Datasets and data collection}

Our study relies on six Telegram datasets originating from three main data sources, each differing in collection methodology and linguistic scope. In particular, two of these sources were previously published in the literature~\cite{baumgartner2020pushshift, alvisi2025mapping}, while the third was collected specifically for this work, as described in the following.
\begin{itemize}
    \item \textit{Dataset from Alvisi et al.\cite{alvisi2025mapping}}:
    This dataset contains more than 186 million messages posted in Italian Telegram chats between January 1 and December 31, 2023. It was collected with an iterative snowball approach that began from a set of seed channels and expanded by following forwarded messages. We refer to it as \texttt{Italian2}.
    
    \item \textit{Dataset from Baumgartner et al.\cite{baumgartner2020pushshift}}:
    This large-scale Telegram dataset includes 96 million chats and 317 million messages collected between 2015 and 2019 using a snowball strategy. Since it includes messages from multiple languages and topics, we extracted four subsets for cross-linguistic comparison, which we refer to as \texttt{English1}, \texttt{Italian1}, \texttt{Russian}, and \texttt{Portuguese}.
    
    \item \textit{Original dataset collected for this study}: 
    Our third dataset is an English-language Telegram dataset collected specifically for this study. Starting from the list of English Telegram channels available on tgstat.com, we retrieved all associated public group messages posted in 2024. This dataset differs slightly from the previous ones, as we did not employ an iterative snowball crawling approach. It contains more than 238 million messages, as detailed in Table~\ref{tab:datasets_ccdf}. We refer to it as \texttt{English2}.
\end{itemize}

Overall, these datasets contain more than 522 million messages. For consistency, we refer to all language-specific subsets as separate datasets, as shown in Table~\ref{tab:datasets_ccdf}. 

\begin{table}[t]
    \centering
        \begin{tabular}{lcrrr}
        \toprule
        \textbf{Dataset} & \textbf{Time span}  & \textbf{Chats} & \textbf{Users} & \textbf{Messages}   \\
        \midrule
        English1\cite{baumgartner2020pushshift} & 2015--2019 & 24,503 & 431,214  & 29,447,990\\
        Russian\cite{baumgartner2020pushshift}& 2015--2019& 11,346 & 342,619  & 48,569,460  \\
        Italian1\cite{baumgartner2020pushshift}  & 2015--2019& 6,536  & 103,892  & 8,928,252 \\
        Portuguese\cite{baumgartner2020pushshift} & 2015--2019   & 6,592  & 96,467  & 9,732,694\\
        Italian2\cite{alvisi2025mapping} & 2023 & 15,871 & 1,307,169  & 186,757,857  \\
        English2 & 2024 & 12,074 & 9,691,170  & 238,923,774 \\
        \midrule
        \textbf{Total} & \textbf{-} & \textbf{-} & \textbf{-} & \textbf{522,360,027}\\
        \bottomrule
    \end{tabular}

    \caption{Overview of the datasets used in our analyses.}
    \label{tab:datasets_ccdf}
\end{table}

\subsection*{Data preprocessing}
To ensure the reliability of our analyses, we applied a multi-stage data processing pipeline designed to guarantee comparability across datasets and languages. First, we restricted the corpus to user-generated textual messages only, excluding system notifications, bot messages, automatic service messages, and any form of non-textual media (e.g., images, videos, stickers, GIFs, voice notes). Since Perspective API cannot label non-linguistic content or very short tokens, we additionally removed empty messages, messages containing only URLs, emojis, or non-recognized characters, as well as messages that the API failed to classify.

Given the centrality of user-level behavioral comparisons across multiple communities, we imposed strict filtering criteria on user activity. Specifically, we included only those users who produced at least one hundred messages within a given chat and who satisfied this activity threshold in at least two distinct chats. This constraint ensured that the users retained for analysis were active participants rather than occasional contributors, allowing us to compute stable estimates of their behavior and to evaluate how it varied across different contexts.

After applying these filters, each dataset was transformed into an aggregated structure that captured the interactions between individual users and their corresponding chats, expressed through (chat toxicity, user toxicity) pairs. We summarize the resulting filtered datasets in Table~\ref{tab:data couples}. Overall, this preprocessing pipeline allowed us to focus on users with a sufficiently rich interaction history, ensuring accurate measurement of toxicity patterns and enabling meaningful cross-community comparisons.

\begin{table}[t]
    \centering
        \begin{tabular}{lrrrcc}
        \toprule
        \textbf{Dataset} & \textbf{Active Users} & \textbf{Chats} & \textbf{Toxicity Pairs} & \begin{tabular}[c]{@{}c@{}}\textbf{Avg. Chat Toxicity}\\ \textbf{($\mu \pm \sigma$)}\end{tabular} & \begin{tabular}[c]{@{}c@{}}\textbf{Avg. User Toxicity}\\ \textbf{($\mu \pm \sigma$)}\end{tabular} \\
        \midrule
        English1     & $1,740$     & $465$   & $4,607$     & $2.45 \pm 2.70$ & $4.15 \pm 4.54$ \\
        Russian      & $3,202$     & $544$   & $7,630$     & $1.36 \pm 1.52$ & $2.00 \pm 4.54$ \\
        Italian1     & $461$       & $88$    & $1,041$     & $1.69 \pm 1.47$ & $1.45 \pm 1.85$ \\
        Portuguese   & $646$       & $86$    & $1,578$     & $1.26 \pm 1.31$ & $1.62 \pm 2.32$ \\
        Italian2            & $20,875$    & $3,679$ & $67,902$    & $1.34 \pm 1.63$ & $1.62 \pm 3.48$ \\
        English2                                    & $11,515$    & $1,540$ & $32,960$    & $1.47 \pm 2.95$ & $1.70 \pm 4.87$ \\
        \bottomrule
    \end{tabular}
    \caption{\label{tab:data couples}
    Aggregated dataset after preprocessing. Only users with at least $100$ messages in two or more chats are included (i.e., active users). Each toxicity pair corresponds to a (chat--toxicity, user--toxicity) combination.}
\end{table}

\subsection*{Correlation measures between user and chat toxicity} 
As the datasets' distributions span several orders of magnitude and displays different tail behaviors, heavy-tailed or light-tailed, as shown in Figure~\ref{fig:ccdf}, we employed logarithmic binning for visualization and analysis. Logarithmic binning reduces noise in sparsely populated regions of the distribution, while also stabilizing the estimates in the tails. Although it is classically applied to heavy-tailed data\cite{clauset2009power,milojevic2010power}, it is also appropriate for light-tailed or rapidly decaying distributions, as it provides a balanced representation~\cite{virkar2012power}. 

Each dataset was divided into 40 logarithmically spaced bins, and bins with fewer than 10 observations were discarded to avoid undersampling. We then computed the Pearson correlations using the midpoint of the bin borders and the mean value of the elements in each bin. We used the \texttt{pearsonr} method of the \texttt{scipy} package to compute both the correlation values, as well as the p-values.
Moreover, to ensure that our results did not depend on the specific binning procedure, we repeated all analyses, varying the number of bins from 20 to 100 in steps of 10, and the minimum bin size threshold from 10 to 20 in steps of 2. A recap of this validation, for each analysis, can be found in Table~\ref{tab:validation}. This table shows that the median values obtained from all validation runs for each analysis remain consistently high and comparable to the original results. In addition, the standard deviation is relatively low, indicating stable outcomes.

\begin{table}[t]
\centering
\begin{tabular}{lcc}
\toprule
\textbf{Dataset} & 
\begin{tabular}[c]{@{}c@{}}\textbf{Static (Figure~\ref{fig:q1}a)}\end{tabular} & 
\begin{tabular}[c]{@{}c@{}}\textbf{Dynamic (Figure~\ref{fig:delta})}\end{tabular} \\

\midrule
English1   & $0.995 \pm 0.002$ & $0.863 \pm 0.065$ \\
Russian    & $0.979 \pm 0.002$ & $0.800 \pm 0.106$ \\
Italian1   & $0.982 \pm 0.007$ & $0.882 \pm 0.071$ \\
Portuguese & $0.989 \pm 0.002$ & $0.766 \pm 0.073$ \\
Italian2   & $0.992 \pm 0.007$ & $0.908 \pm 0.030$ \\
English2   & $0.997 \pm 0.002$ & $0.990 \pm 0.003$ \\
\bottomrule
\end{tabular}
\caption{Robustness to binning choices.
Median Pearson correlations ($\mu$) with standard deviation ($\sigma$) across validation runs
varying the number of logarithmic bins (20--100) and the minimum bin size threshold (10--20, step~2).
Columns correspond to the same analyses shown in Figures~\ref{fig:q1}a and~\ref{fig:delta}, confirming that correlations remain stable
across binning configurations.}
\label{tab:validation}
\end{table}


\subsection*{Fitting user conformity patterns}

To approximate the behavior of individual users, we fitted a linear model relating the toxicity of each user in a chat to the overall toxicity of that chat. This was done across all chats in which the user participated, using the \texttt{LinearRegression} module from the \texttt{scikit-learn} package~\cite{scikit-learn}. The resulting slope and intercept of the fitted line provide a characterization of the user's behavior, as shown in Figure~\ref{fig:q2}a. We conceptualize this slope as a conformity index, where $\theta$ denotes the regression slope expressed in angular degrees. 
We the define the slope $\theta$ as the inverse of the tangent of the angular coefficient, and thus we classify users according to the direction and magnitude of their slope:
\begin{itemize}
    \item \textit{Conformist users}: $\theta > \tau$ (i.e., toxicity increases as the environment becomes more toxic).
    \item \textit{Anti-conformist users}: $\theta < -\tau$ (i.e., toxicity decreases as the environment becomes more toxic).
    \item \textit{Independent users}: $|\theta| \leq \tau$ (i.e., toxicity remains stable across contexts).
    \item \textit{Zen users}: users who never displayed any toxicity in any chat.
\end{itemize}

A key methodological challenge is defining what counts as ``near to zero'' when distinguishing independent users from conformists or anti-conformists. Since raw regression coefficients lack an intuitive scale, we convert slope values into angular degrees, where $0^\circ$ corresponds to a flat line. This allows us to interpret thresholds as angular deviations from behavioral stability.
In the main analysis, we set $\tau=15$: users whose slope lies between $-15^\circ$ and $+15^\circ$ are considered independent, while slopes exceeding that range indicate conformity or anti-conformity. This cutoff offers a reasonable trade-off between sensitivity and interpretability.

\begin{figure}[t]
  \centering
  \includegraphics[width=\textwidth]{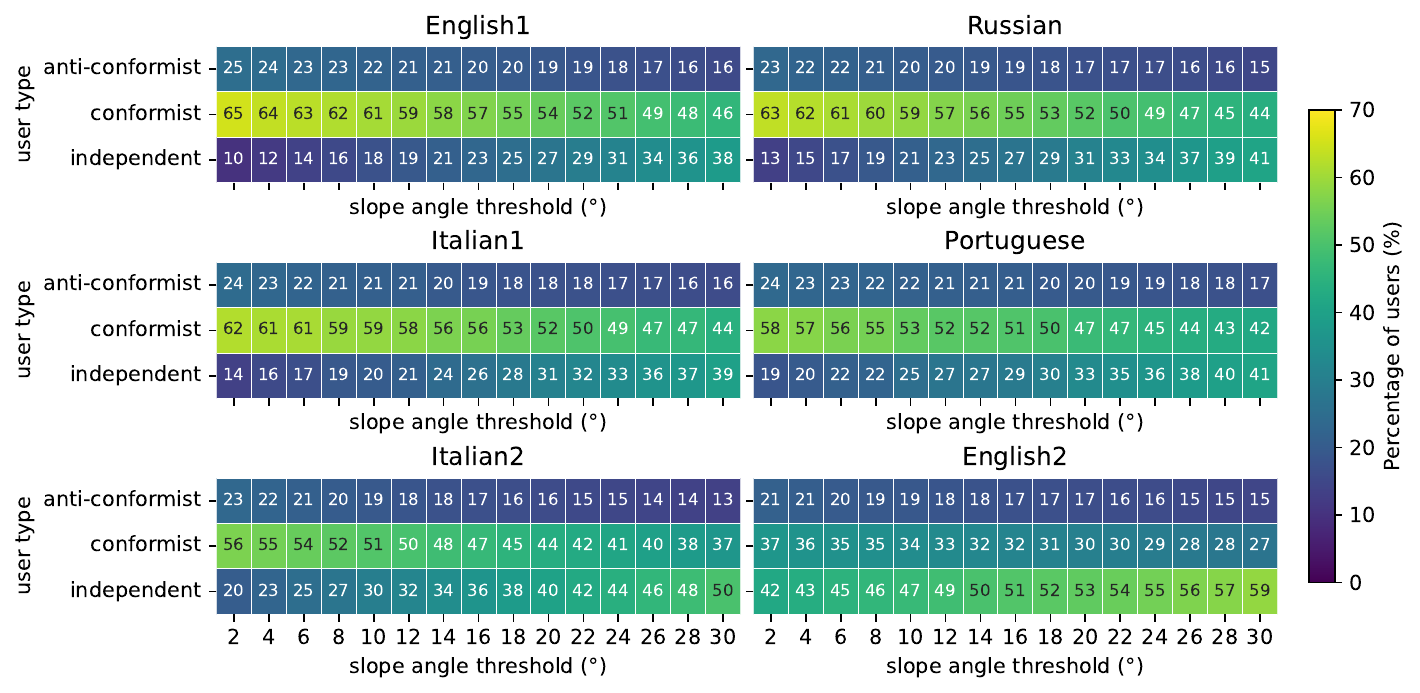}
  \caption{Sensitivity analysis of user classification as a function of the slope angle threshold $\tau$. The plot shows how the proportions of conformist, independent, and anti-conformist users vary as the threshold used to define independence is adjusted from $2^\circ$ to $30^\circ$. Users are classified as independent when $|\theta| < \tau$, where $\theta$ is the regression slope expressed in angular degrees. Despite changes in the threshold, the ratio between conformists and anti-conformists remains approximately stable, indicating that the predominance of conformist behavior is robust across a broad range of cutoff values.}
  \label{fig:thresh-origin-label}
\end{figure}

To assess the robustness of our findings, we systematically vary the cutoff used to determine whether a user's slope is sufficiently close to zero to be considered independent. Specifically, we test a wide range of angular thresholds, treating users as independent whenever their absolute slope angle satisfies $|\theta| < \tau$, with $\tau$ ranging from $2^\circ$ to $30^\circ$ (Figure~\ref{fig:thresh-origin-label}). Across the entire threshold range, the relative proportions of conformist, independent, and anti-conformist users remain qualitatively stable, indicating that the prevalence of conformist behavior is not an artifact of a particular cutoff choice and that the relative balance between conformist and ant-conformists remains stable across datasets.

\bibliography{sample}

\section*{Author contributions statement}

L.A, V.P., G.C., S.T. and M.T. conceived the study and designed the research framework. L.A. and V.P collected the data and preprocessed the data. V.P., L.A., G.C., S.T. developed the computational models and performed the analyses. V.P., L.A, G.C., S.T. and M.T. contributed to the interpretation of the results. V.P., L.A, G.C., S.T. and M.T.  wrote the initial draft of the manuscript. All authors discussed the results, contributed to the final version of the manuscript, and approved it for submission.

\section*{Additional information}

To include, in this order: \textbf{Accession codes} (where applicable); \textbf{Competing interests} (mandatory statement). 

The corresponding author is responsible for submitting a competing interests statement on behalf of all authors of the paper. This statement must be included in the submitted article file.
\end{document}